\documentclass[submission,copyright,creativecommons]{eptcs}
\usepackage{breakurl}             
\usepackage{graphicx}
\graphicspath{ {graphics/} }

\newcommand{\csdv}{CloudSDV}
\newcommand{\sdv}{SDV}
\newcommand{\smv}{SMV}

\title{Static Analysis Using the Cloud}
\author{ Rahul Kumar
\institute{Microsoft Research, Redmond, WA, USA}
\and
Chetan Bansal
\institute{Microsoft Research, Redmond, WA, USA}
\and
Jakob Lichtenberg
\institute{Microsoft, Redmond, WA, USA}
}

\def\titlerunning{Static Analysis Using the Cloud}
\def\authorrunning{Kumar et al.}

\begin{document}
\maketitle

\begin{abstract}
In this paper we describe our experience of using Microsoft Azure
cloud computing platform for static analysis. We start by extending
Static Driver Verifier to operate in the Microsoft Azure cloud with
significant improvements in performance and scalability. We present
our results of using \sdv{} on single drivers and driver suites using
various configurations of the cloud relative to a local
machine. Finally, we describe the Static Module Verifier platform, a
highly extensible and configurable platform for static analysis of
generic modules, where we have integrated support for verification
using a cloud services provider (Microsoft Azure in this case).
\end{abstract}

\section{Introduction}

The last decade has seen a marked increase in the use of formal
methods and static analysis in a variety of domains such as software
development and systems engineering. Applications of formal methods
vary from defect discovery to automated/manual theorem proving for
performing analysis and proving system correctness. The Static Driver
Verifier tool is one such static analysis tool that enables the
discovery of defects in Windows device
drivers~\cite{ball2004slam,msSDV}. It is and has been used with great
effectiveness to check Windows device drivers for API
compliance~\cite{msDDI}. Currently, \sdv{} is shipped as part of the
Windows Driver Development Kit. As with other program analysis and
defect discovery tools, the major issues with \sdv{} are related to
the {\em performance} and {\em scalability} of the tool. By {\em
  performance}, we refer to the total amount of time and memory
resources required for performing the verification task (e.g., wall
clock time, memory pressure etc.). {\em Scalability} on the other
hand, is the {\em size of the driver} that we are able to successfully
verify by proving the absence or presence of defects with no concern
for utilized resources. Small drivers may be completely verifiable by
\sdv{}, but still face a performance problem because it takes a long
time to completely verify them. For example, for WDM drivers, there
are approximately 200 rules that need to be verified for each device
driver. The maximum time that each rule can take to be verified is 50
minutes; thus, resulting in a maximum possible total of 10000 minutes,
or 167 hours. Even on a multi-core computer, the verification can
possibly span days. For device driver developers, this can be
extremely frustrating and negative, possibly resulting in finally not
using the tool for verification and losing confidence in static
analysis in general. It is important to note that this particular
problem can easily get worse if more rules are developed and verified
on device drivers. Past experiences of using \sdv{} on larger device
drivers proved valuable for understanding the scalability problem that
\sdv{} runs into.

We attempt to solve some of these problems by enabling \sdv{} to use a
cloud platform such as Microsoft Azure~\cite{azure}. \csdv{} is a
Microsoft Azure based computation system that allows \sdv{} to farm
out its verification task to the cloud. Doing so, provides benefits in
multiple different areas for both, users as well as developers of
\sdv:

\begin{itemize}

\item {\em Parallelize.} Multiple verification tasks can be dispatched
  simultaneously for parallel computation; thus, improving the
  performance of the verification run.

\item {\em Offline Computation.} By farming out verification tasks to
  the cloud, it is possible to schedule the entire verification of a
  driver and re-visit the results at a later point.

\item {\em Result storage.} Using cloud infrastructure, it is now
  possible to perform better result storage and telemetry for
  \sdv{}. Results that were previously produced on local machines, are
  now recorded systematically in the cloud for future analysis
  (telemetry, learning etc.). Additionally, it also becomes easier to
  {\em query} for results in the past, whereas previously, such
  results would have been lost permanently.

\item {\em Verification as a Service.}  Similar to other software
  services, \csdv{} allows improvements and bug fixes to \sdv{} to be
  distributed much more easily. Updates can also be distributed with a
  much higher cadence as opposed to being governed by a less frequent
  schedule of the parent software (in this case the Windows Driver
  Development Kit).

\item {\em Scalability.} Moving verification tasks to the cloud does
  not solve the scalability problem in the traditional sense, but does
  provide the opportunity to verify larger drivers that have been
  resource intense to verify until now. For example, allowing a much
  larger timeout for each rule does not significantly increase the
  total verification time anymore. Rather, only a much smaller penalty
  is associated with the larger timeout; thus, allowing users to
  potentially get better results.

\end{itemize}

Based on our experience with \sdv{} and other static analysis tools, we designed Static Module Verifier (\smv{}), a generic and extensible static analysis platform with integrated cloud support. \smv{} includes all the essential components of static analysis tools such as build interception, extensibility, storage and cloud integration.

The paper is organized as follows. Section~\ref{sec:sdv} gives a brief
introduction to \sdv{}. Section~\ref{sec:design} gives an overview of
the architecture and implementation of
\csdv{}. Section~\ref{sec:results} presents the results of using
\csdv{} on various drivers and test suites. Section~\ref{sec:smv} talks about the motivation and the goals behind \smv{}, while in Section~\ref{sec:architecture} we discuss it's architecture. Section~\ref{sec:related}
presents a brief overview of known related work in this area and
Section~\ref{sec:conclusions} presents a discussion and conclusions.

\section{\sdv{} Background} 
\label{sec:sdv}

\begin{figure*}
\centering
\includegraphics[scale=0.4]{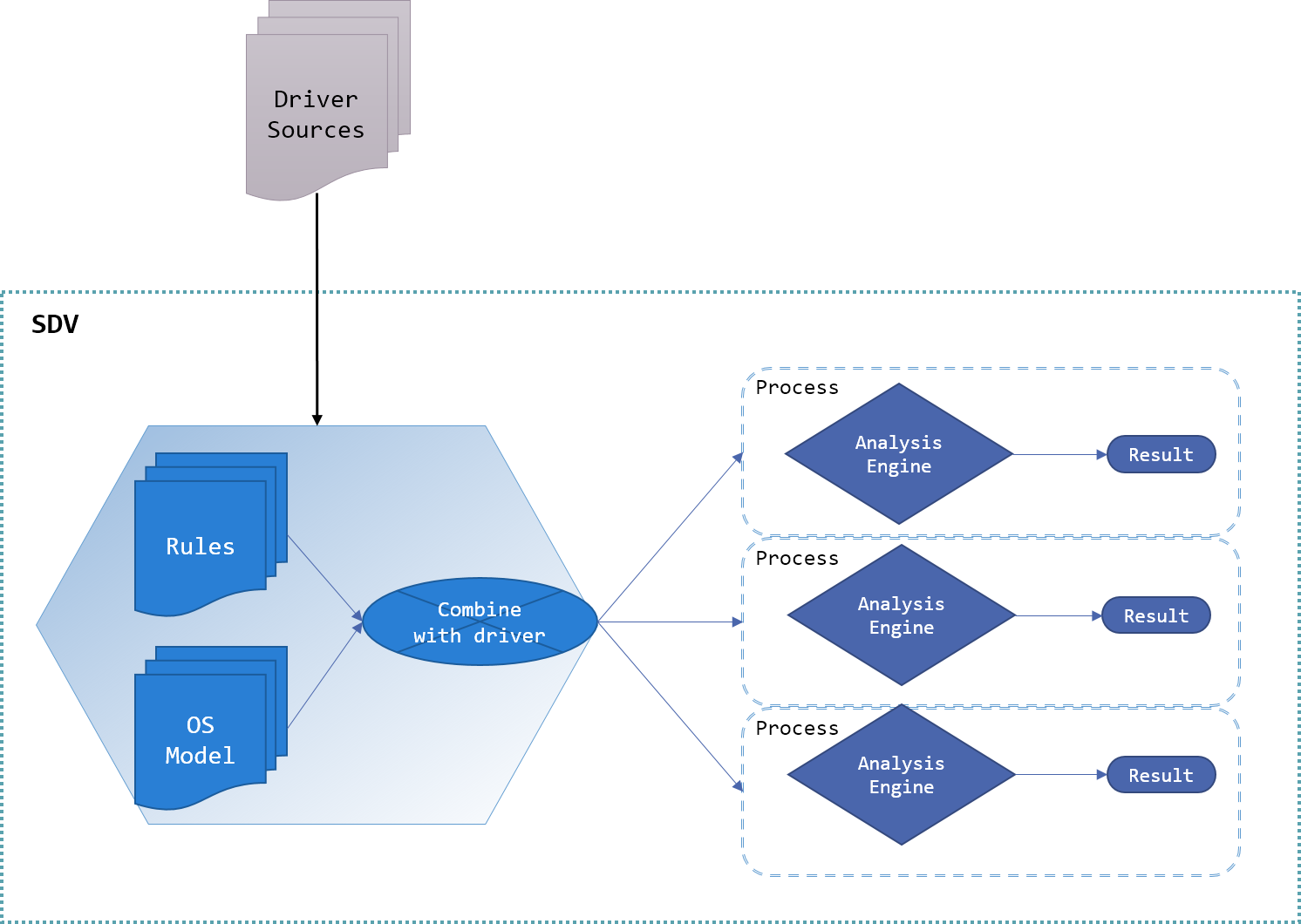}
\caption{An overview of the \sdv{} system.
\label{fig:sdv}}
\end{figure*}

We first present a brief overview of \sdv{} and how it
functions. Figure~\ref{fig:sdv} shows a high level overview of how
\sdv{} operates. 

\sdv{} takes as input the sources of the driver under test and
optionally, a set of rules that are to be verified on the
driver. Given a driver, \sdv{} first ensures that the driver can be
compiled correctly. Following this initial compilation, \sdv{} then
proceeds to combine the driver sources with the Operating System model
and the {\em rules} that are to be verified on the driver. The
Operating System model inside \sdv{} is a set of C source files that
model the Windows operating system in a manner that is most useful for
performing verification by the analysis engine within \sdv{}. There
are two primary parts to the OS model. The {\em stubs} that provide a
demonic model of the kernel APIs that can possibly be called by
Windows device drivers, and the {\em harness} that simulates the
operating system and how it would call into the driver to perform
functions. The {\em rules} within \sdv{} express temporal safety
properties that drivers must adhere to. These rules are expressed in
SLIC, which is a C-like language streamlined to express safety
automaton~\cite{ball2002slic}. The result of combining the driver
sources with the operating system model and rules is an intermediate
binary file that can be consumed by the analysis engine for
verification. It should be noted that each rule that is to be verified
produces a new unique intermediate file. Thus, for N rules, there are
N unique binary files that are created for performing the
verification.

Until now \sdv{} has only prepared the work that needs to be performed
for verifying the driver. Once the verification files are ready for
all the rules, \sdv{} schedules the verification tasks in parallel,
limiting the number of concurrent verification tasks to the total
number of logical cores available in the local machine. Each
verification task is limited to a {\em timeout} (in seconds) and {\em
  spaceout} (in MB) value that can be specified as a configuration to
\sdv{}. \sdv{} waits for each verification run to complete and
eventually reports results back to the user. Until the \csdv{} work
presented in this paper, \sdv{} was only capable of running on a
single machine. For performing mass scale verification of many
drivers, additional infrastructure had been developed for executing
\sdv{} in parallel on multiple machines, but such infrastructure was
unintuitive and extremely tedious to use.

\section{\csdv{} Design and Implementation}
\label{sec:design}

We now describe the architecture, design, and implementation details
of \csdv. \csdv{} is implemented using C\# and the .NET Microsoft
Azure API. The analysis engines and other parts of \sdv{} have been
implemented using OCaml, C, and C++.

\begin{figure*}
\centering
\includegraphics[scale=0.4]{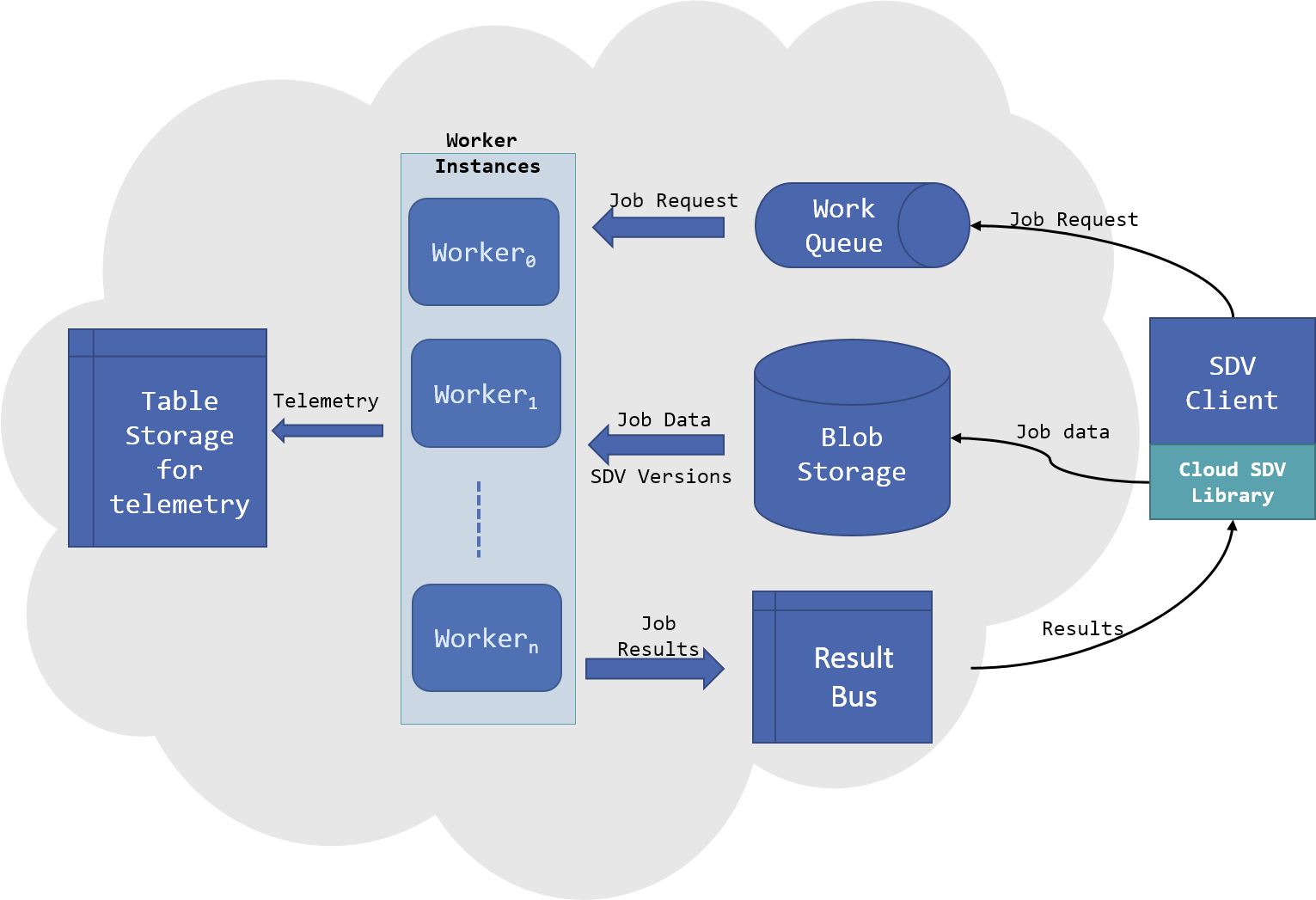}
\caption{An overview of the \csdv{} system.
\label{fig:csdv}}
\end{figure*}

\begin{figure}
\centering
\includegraphics[scale=0.4]{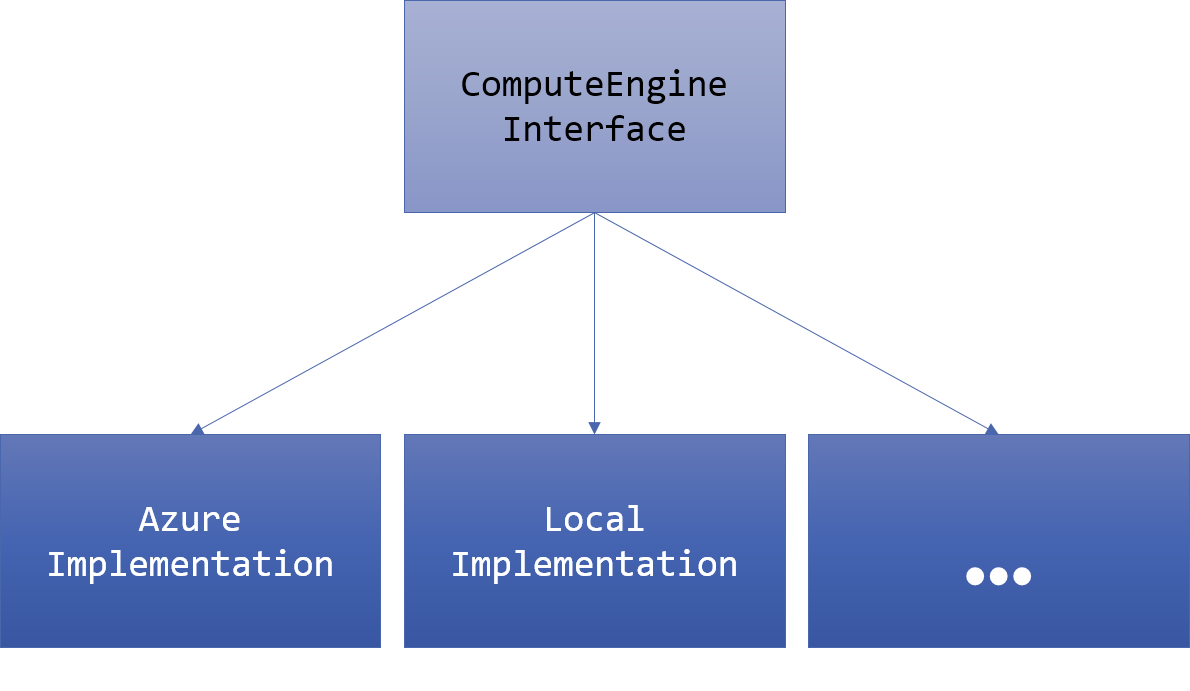}
\caption{Various implementations of the compute engine interface.
\label{fig:interface}}
\end{figure}

Figure~\ref{fig:csdv} shows a high level overview of the \csdv{}
system.  In general, time flows from right to left (counter clockwise)
in the figure. The right most side of the figure contains the \csdv{}
client, which is slightly different from the normal \sdv{} client
described earlier. There is an extra option included in the normal
\sdv{} client to enable the \csdv{} scenarios. The \csdv{} client can
be partitioned into two distinct parts. The first is the \csdv{}
library that is a generic library that allows any application to
interact with the \csdv{} Microsoft Azure service, and the second, is
the \sdv{} client slightly modified to interact with the \csdv{}
library. The \csdv{} library provides a generic interface (the compute
engine interface) that allows an application to schedule a
verification task, upload files and other data for performing a
verification task, and retrieve results from the computation
platform. Figure~\ref{fig:interface} illustrates the current
implementations of the interface, with the bottom right hand box
representing future implementations for other parallelization
platforms. Based on the configuration of the application or command
line arguments, the \csdv{} client can either choose to perform the
verification tasks locally, or using the Microsoft Azure computation
platform. In the future, we plan on providing additional
implementations that can take advantage of other computation platforms
such as clusters, super computers etc. It should be noted that
currently, in all cases, the given driver is always compiled on the
local machine and all the verification tasks themselves are created on
the local machine.

After the \csdv{} client has compiled the driver and produced the
various verification tasks that need to be run, each task is
scheduled. This translates to two actions for each verification task:
uploading the relevant payload/data to the cloud computation platform,
and scheduling the verification task by inserting an entry into the
work request queue. The payload/data is uploaded to Microsoft Azure's
blob storage~\cite{azureBlobStorage}, which allows storing large
amounts of unstructured text or binary data for random direct
access. The Azure queuing service is used for storing verification
task requests~\cite{azureQueueStorage}. Each verification task in the
queue contains pointers to the data that has been previously uploaded
by the \csdv{} client. Within the \csdv{} Azure implementation, there
exist \csdv{} worker instances (grouped vertically in the middle of
the cloud). The worker instances are responsible for polling the work
queue for new verification tasks that have been submitted by \csdv{}
clients. 

\begin{figure}
\hspace{1.5cm} \includegraphics[scale=0.40]{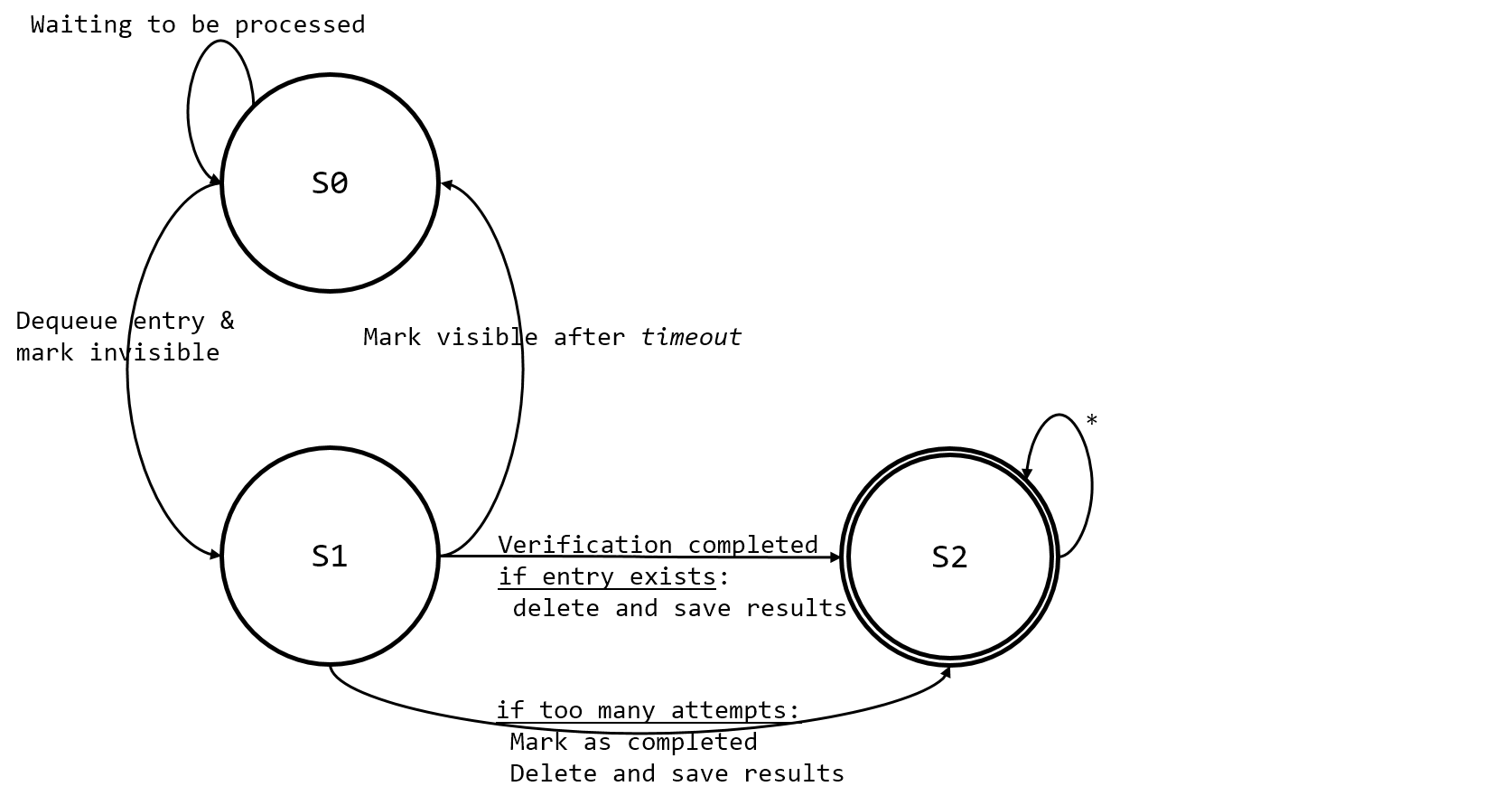}
\caption{State machine of the queuing system employed by \csdv{}.
\label{fig:queue}}
\end{figure}

Figure~\ref{fig:queue} shows the state machine implemented by \csdv{}
for dealing with messages in the request queue. As soon as a worker
instance discovers a new task in the queue, the worker instances marks
the task (queue item) as {\em invisible} to other instances. This is
done {\em atomically}, so that other worker instances are unable to
retrieve the same message. The transition from states $S0$ to $S1$
reflects this behavior. In state $S1$, the worker is processing the
verification task specified in the queue entry. Azure queues provide
the ability to mark the message invisible for a specific time
period. For a given {\em timeout} value specified to \csdv{}, we use
$timeout + 300$ seconds as the invisibility period of the entry in the
queue. After that time limit has been reached, the entry is
automatically marked visible again. This is represented as the
transition from states $S1$ to $S0$. This transition is {\em only}
executed in cases where the worker was unable to process the task
successfully within the allocated amount of time (due to unknown
reasons arising from faulty code or unexpected crashes). In such
cases, the result of the worker can be considered lost, and having
marked the task as visible allows us to guarantee processing it
again. In the case when the worker successfully processes the
verification task, the entry is deleted from the queue permanently, as
shown by the transition from states $S1$ to $S2$. State $S2$ is a
final state where the entry no longer exists in the queue. We choose
$timeout + 300$ seconds as the value of the invisibility period to
allow for delays. For example, it may be possible that the
verification task completes extremely close to the $timeout$ period,
in which case, in highly congested network conditions, the delete
message from the worker to the Azure queuing system may only arrive
after the $timeout$ has expired. The additional 300 second (arbitrary
choice) buffer allows us to account for this scenario. Even with the
additional buffer time, under extreme circumstances it is possible
that the entry may be marked visible and then later deleted from the
queue by the worker that originally dequeued the entry. In such a
case, there is only a side effect if another worker has dequeued the
same entry and marked it invisible before the original worker's delete
message was processed, but after the Azure queuing system marked the
entry as visible. In this particular scenario, the entry is deleted by
the original worker and the second worker (after performing the
verification) eventually discards its results and proceeds to the next
work item in the queue (based on an existence check in the queue). The
disadvantage in this scheme is the possibility for redundant
processing (at most one extra time), which arises in part due to
Azure's at-least-once best effort FIFO queue semantics. In our
experience and experiments, we never encountered this situation, which
we believe is in part due to the large buffer we specify in the
invisibility timeout period. On the other hand, the advantage of
marking items as invisible/visible in the queue is the strong
guarantee of always processing all tasks in the queue. Finally, we
also impose a limit on the number of times an entry can be dequeued
from the queue. Each time an entry is dequeued for processing, we
modify the entry and increment the dequeue count. Once the specified
limit is reached, the entry is deleted from the queue and the result
of that verification task is listed as a \texttt{ToolError}. This
condition is implemented to account for situations where the tools
encounter an irrecoverable deterministic error and workers repeatedly
try to complete the same task.

Both the $timeout$ and the $spaceout$ values are used to limit the
amount of time and memory that can be consumed by the verification
task on the worker instance. If these values are exceeded, the
verification task is ended and the result is marked as a
\texttt{TimeOut} or \texttt{SpaceOut}. These limits are user
configurable only on the \csdv{} client. Once a verification task has
been submitted to the queue, the configuration of the task can no
longer be modified. The space and time restrictions for the
verification task are enforced using a monitoring agent on the worker
instance.

For each task that is executed, the instance also inserts telemetry
data into Azure table storage~\cite{azureTableStorage}. Telemetry data
is used for performing analysis of \csdv. Telemetry data and analysis
includes various metrics such as time taken to complete task, average
time a task was waiting in the queue before retrieval by a worker,
number of times the task was marked visible/invisible etc. Telemetry
data is anonymous and used solely for the purpose of studying the
system to improve performance, stability, and efficiency.

During the entire time the verification is occurring in the cloud, the
\csdv{} client is continuously polling the Result service bus for new
available results. The service bus is a commonly accessible storage
medium (between workers and \csdv{} client) where workers publish new
results to a {\em topic}. Each topic in the result bus is the unique
ID of the \csdv{} client that submitted verification tasks. This ID is
made part of the task details, so the workers know where to publish
the result. As soon as results are available for any scheduled tasks,
the \csdv{} client reports the results to the user and exits. It
should be noted that there is extremely little difference (regarding
GUI and console output) in the experience the end user gets when using
\csdv{} relative to just using the \sdv{} client, which is considered
a positive aspect of the \csdv{} system.

\subsection{\csdv{} and \sdv{} Versions}

A salient feature provided by \csdv{} is the ability to use different
versions of the core \sdv{} product. This feature is especially useful
for development and testing of new \sdv{} versions, where one is
interested in comparing two versions of \sdv{} against each other. The
feature enables \sdv{} developers (analysis engine, rule, operating
system model) to upload their custom versions of \sdv{} to the Azure
blob storage and run experiments on a mass scale very quickly. Using
existing infrastructure, the developers are then able to regress their
current results against saved baselines.

To enable multiple version support, each component of \csdv{} must be
aware of the specific version being used. The \csdv{} client can be
configured to specify the \sdv{} version that is to be used for the
verification. Each verification task that is created from that client,
will also contain the same version string that is to be used for
verification. In the Azure cloud, when each worker instance starts
processing a new task request, it first checks to see if the \sdv{}
version specified is present on the worker instance or not. If
present, the worker instance switches to using the specified \sdv{}
version and completes the task. If not present, the worker instance
first downloads the specified \sdv{} version from \csdv's private blob
store. If the version cannot be found in the blob store, the worker
instance marks the verification task as completed and provides an
error code as the final result.

\sdv{} developers (infrastructure, OS model, analysis engines, and
rules) are provided special access and instructions for uploading
their private versions of \sdv{}. 

\subsection{\csdv{} Monitor}

As part of the \csdv{} infrastructure, we also implemented a simple
\csdv{} monitoring tool that allows us to monitor the current state of
a \csdv{} deployment. Our goal is to have the monitor serve as a one
stop location for administrators to view/modify the status of
\csdv{}. The monitor is deployed as a Microsoft Azure application
itself. As configuration, it takes the list of \csdv{} deployments to
monitor. Currently we have two deployments in Asia and US. Given a
deployment, the \csdv{} Monitor shows basic information about the
deployment, the number of current active workers, and the status of
the task queue. For each entry, we display the globally unique
identifier for the task, the name of the driver, the rule being
verified on the driver, the version of \sdv{} requested the submission
time, and the exact command that is to be used for performing the
verification task. Future versions of the monitor are intended to be
more interactive and enable administrators to perform operations on
individual tasks or the entire deployment itself. It is also possible,
that in future versions, users of the cloud verification service will
be able to submit, cancel, and monitor jobs.

\subsection{Autoscale}

To minimize costs incurred on a regular basis, the \csdv{} system
always operates with a default of two worker instances. Along with the
default 2 workers, we use the Autoscale~\cite{azureAutoscale} feature
of Microsoft Azure to adjust to incoming verification tasks. The
Autoscale feature is configured to increase the number of active
workers based on the size of the queue. Since Azure monitors the
length of the queue at fixed intervals (\csdv{} uses an interval of 10
minutes), any time the length of the queue exceeds the configured
limit, a certain number of new instances of the workers are created
(\csdv{} specifies this number as 50). This process can potentially
repeat itself until an upper limit of the total number of worker
instances is reached (for \csdv{} this limit is 200). It should be
noted that the new worker instances created are not immediately
available; it can take anywhere up to 10 minutes for the workers to be
fully functional and available, although, in practice we observe that
all the new worker instances are available between 2 and 3
minutes. Further, each worker once activated has no \sdv{} versions
available to it. Rather, a new worker lazily acquires \sdv{} versions
as new tasks from the queue are processed. All results presented in
this paper are using the Autoscale feature as described here. We
believe it would be interesting to repeat the experiments with
different configurations of the Autoscale feature, or to disable
Autoscale and always have 200 instances available. We plan on doing
such experiments in the future.

\subsection{Results}
\label{sec:results}

\begin{figure}
  \centering
  \begin{tabular}[c]{c}
      \includegraphics[scale=0.4]{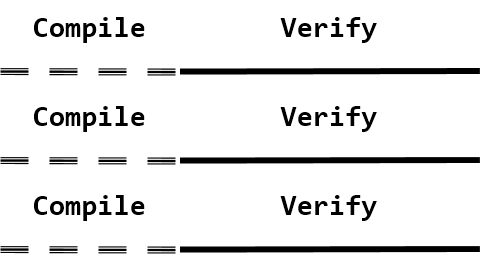}      
      \vspace{0.3cm} 
      \\ 
      (a) 
      \vspace{0.3cm} 
      \\
      \includegraphics[scale=0.5]{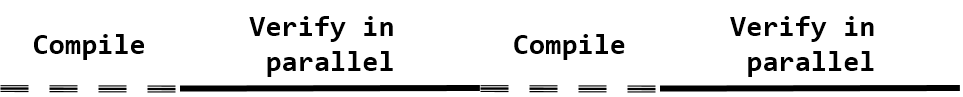} 
      \vspace{0.3cm}
      \\ 
      (b) 
      \\
  \end{tabular}    
  \caption{A visualization of the parallelism schemes implemented in
    (a) \csdv{} and (b) the \sdv{} test infrastructure.}
  \label{fig:parallel}
\end{figure}

We now present our test methodology and results of using the \csdv{}
system. To test \csdv, we first selected two drivers that vary in size
and complexity. These drivers have been a part of the \sdv{} test
suite for a long time and continue to serve as good baseline drivers
for testing. \texttt{fail\_driver1} is an extremely small and simple
driver that \sdv{} can verify relatively easily. The \texttt{serial}
driver is a much larger driver that takes a very long time to compile
and verify. An 8 core machine may take 6 hours to compile and verify
all 200 rules. After testing on individual drivers, we tested \csdv{}
using test suites that are also available in \sdv{}. The test suites
specify a set of drivers and rules to be checked on the drivers. Each
rule results in a single verification task, so in total, there are
$drivers \times rules$ checks that are produced. Typically, the test
suites can be run locally, or using the \sdv{} test infrastructure
which will distribute the tasks over a set of machines (managed
personally). It should be noted that \sdv{} test infrastructure does
not parallelize over all drivers and all checks, rather only over all
checks, one driver at a time. This choice was made to accommodate for
infrastructure limitations. In contrast, when we utilize \csdv{}, we
parallelize over all drivers and all checks. This is possible because
the infrastructure limitations are not relevant anymore. We use the
JOM~\cite{jom} tool to help us parallelize over all
drivers. Figure~\ref{fig:parallel}(a) illustrates the parallelism as
implemented with \csdv{}, and Figure~\ref{fig:parallel}(b) illustrates
the parallelism as implemented by the \sdv{} test infrastructure. Each
dashed line followed by a solid line is one driver being compiled and
then verified.

Table~\ref{tab:results} shows the results of using \csdv{} on various
drivers and test suites. The Local run is performed using an Intel(R)
Xeon(R) CPU with 64 logical cores with a total of 64 GB of memory. The
same machine is also used as the \csdv{} client machine for utilizing
the \csdv{} cloud (for compilations and submitting tasks to the
cloud). For each driver and test suite, we list the total number of
drivers and checks to be performed on the driver along with the time
taken (\texttt{hh:mm}) when using the local machine and when using a
maximum of 20, 100, and 200 worker instances in the \csdv{} cloud.

\begin{table*}
\centering
\begin{tabular}{|l|l|l|l|l|l|l|}
\hline
Driver/Suite & Drivers & Checks & Local & Azure20 & Azure100 & Azure200 \\ 
\hline
\hline
\texttt{fail\_driver1} & 1 & 192 & 00:55 & 00:58 & 00:59 & 00:56 \\
\hline
\texttt{serial} & 1 & 192 & 01:56 & 02:09 & 01:57 & 01:56 \\
\hline
\texttt{sdv\_regress} & 2 & 26 & 00:25 & 00:22 & 00:22 & 00:18 \\
\hline
\texttt{svb-ITP} & 28 & 5040 & 27:26 & 08:11 & 02:20 & 02:08 \\
\hline
\texttt{sdv\_bugbash} & 91 & 16380 & 101:06 & 18:19 & 05:10 & 04:31 \\
\hline
\end{tabular}
\vspace{0.4cm}
\caption{Results of using \csdv{} for various drivers and test suites.
\label{tab:results}
}
\end{table*}

For \texttt{fail\_driver1}, where each verification task is at most 3
seconds (as measured in the past on a local machine), the local
verification run completes in approximately 55 minutes. Most of that
time is spent creating the 192 different verification tasks, 1 for
each rule. Utilizing \csdv{} is actually ineffective in this case
because each verification check has to be transported to the cloud
before it can be completed. The additional overhead of transporting
and waiting for results creates a {\em slowdown}, irrespective of how
many worker machines are utilized. For the \texttt{serial} driver, we
again notice the same behavior, where scheduling tasks on the \csdv{}
cloud results in no significant improvement. Again, this can be
explained by making the observation that the majority of the work
being performed in verifying these drivers is the compilation and
creation of the tasks as opposed to the actual execution of the
verification tasks. The \texttt{sdv\_regress} test suite consists of 2
drivers which take a total of 25 minutes to verify on the local
machine. This is the first time we observe any improvement when using
\csdv{} to verify the drivers. Using 20 workers, we see that the run
only takes 22 minutes, and using 100 and 200 worker instances, the
total time taken is again 22 minutes and 18 minutes
respectively. Since there are only a total of 26 tasks produced by
this suite, we don't observe any significant speedup when going from
20 to 100 workers. For the \texttt{svb-ITP} case, we observe that
there is much more speedup in going from the local run to using 20 and
100 workers. This is because the total number of tasks (5040) produced
is much greater than in any of our experiments before this. Since the
total number of checks is still not significant enough for completely
utilizing the 200 workers, we don't see any significant speedup when
going from 100 to 200 cores. For our last test suite, we picked
\texttt{sdv\_bugbash}, which is by far the largest test suite in terms
of the number of checks it produces. The local run takes more than 4
days to complete. Using even 20 workers produces over a 5x
speedup. Furthermore, \csdv{}'s true value is shown when we move to
100 and 200 cores where we observe speedups of 17.5x and 22.5x
respectively. As seen in the results, the speedup when moving from 100
to 200 cores is modest. This is because all the compilation is still
being performed on a single machine, which does require a significant
amount of time and space resources. We suspect that it is possible to
get even more speedup by using multiple machines for compilation and
adding even more workers.

\begin{table}[t]
\centering
\begin{tabular}{|c|c|c|c|c|}
\hline
Mean & Median & Standard Deviation & Minimum & Maximum\\
\hline
\hline
24.87s & 11.51s & 27.35s & 1s & 91s\\
\hline
\end{tabular}
\vspace{0.5cm}
\caption{Statistics for time spent by a task waiting in the queue.}
\label{tab:qtimes}
\end{table}

Table~\ref{tab:qtimes} shows statistics for time spent by a single
verification task in the queue. These results were gathered over a
total of 3858 checks. As shown in the table, on average, a
verification task spends around 25 seconds waiting in the queue before
a worker starts to process the task. This number can prove to be too
high if the verification task itself is trivial and does not require
much time to be processed and completed. On the other hand, for much
larger tasks that require significantly more time to complete, the
time spent waiting in the queue proves to be trivial and has no
noticeable impact on the entire time taken for verifying the
driver. The maximum time any task spent waiting in the queue was 91
seconds, which is a direct result of the Microsoft Azure Autoscale
feature. This happens when all worker instances are busy with a task
and the tasks in the queue are waiting either for new instances to be
created or for an existing worker instance to poll the queue for a new
task.

\section{\smv{} : Static Module Verifier}
\label{sec:smv}

\begin{figure}
\centering
\includegraphics[scale=0.29]{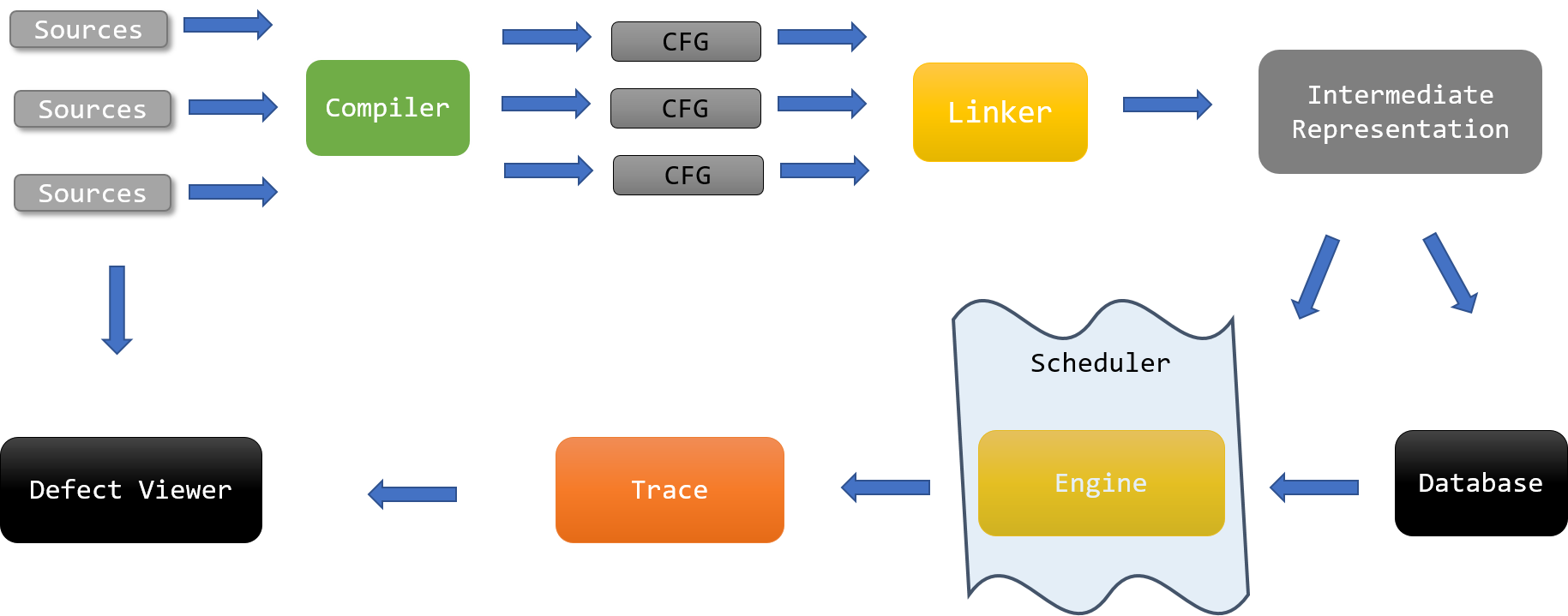}
\caption{Generalized workflow of Static Analysis Tools.
\label{fig:StaticAnalysisWorkflow}}
\end{figure}

Based on our experience of developing and shipping static analysis tools
over the years, we realized that most static analysis tools have
very similar workflows and requirements. As shown in
Figure~\ref{fig:StaticAnalysisWorkflow}, most static analysis tools
require:

\begin{itemize}

  \item {\em Build}. Perform a basic build of the module to verify it
    can be compiled and linked successfully.
  
\item {\em Build Interception}. Build interception to create the
  required build artifacts and Intermediate Representation (IR) of the
  module being analyzed. In most cases this is usually persisting the
  AST or CFG of the input source code, which is further consumable by
  analysis engines. 

\item {\em Portability}. Ability to support multiple build
  environments and platforms. Modules can be built using various
  different type of build systems such as Make~\cite{make},
  Ant~\cite{ant}, MSBuild~\cite{msbuild} etc.

\item {\em Analysis}. Execution of analyses engine on the produced
  build artifacts.

\item {\em Storage}. Storage of build artifacts for future use and re-use .

\item {\em Resources}. Computation and storage infrastructure (such as
  cloud), which can be utilized for processing a large number of tasks
  and storing results.

\end{itemize}

With these objectives in mind, we have created the \smv{}
platform. \smv{} is designed with a plug-in-based architecture that
allows clients (analysis tool developers) easily configure \smv{} for
their environment (build system, compiler etc.) and incorporate their
static analysis tool. Out of the box, \smv{} provides the build
interception technology, the framework for executing the analysis, and
the cloud services. The generated artifacts can also be saved for
future use, and analyses can be run in parallel using Azure cloud
infrastructure. It takes as input the following:

\begin{itemize}
\item Module that is to be built/analyzed.
\item Configuration file for the analysis tool that specifies how the
  module is to be built, how the build is to be intercepted, and what
  analysis if any is to be executed after that.
\item Optionally, a path to the analysis plugin which implements the
  \smv{} plugin interface. This allows the plugin to further customize
  the actions of SMV by performing custom pre and post processing
  between the build, intercepted build, and analysis actions. Examples
  of such customization may include obfuscation of source code for
  sharing defects with third parties, file cleanup, persisting results
  to custom data sinks etc.
\end{itemize}

\section{Architecture}
\label{sec:architecture}
\begin{figure}
  \centering
  \includegraphics[scale=0.33]{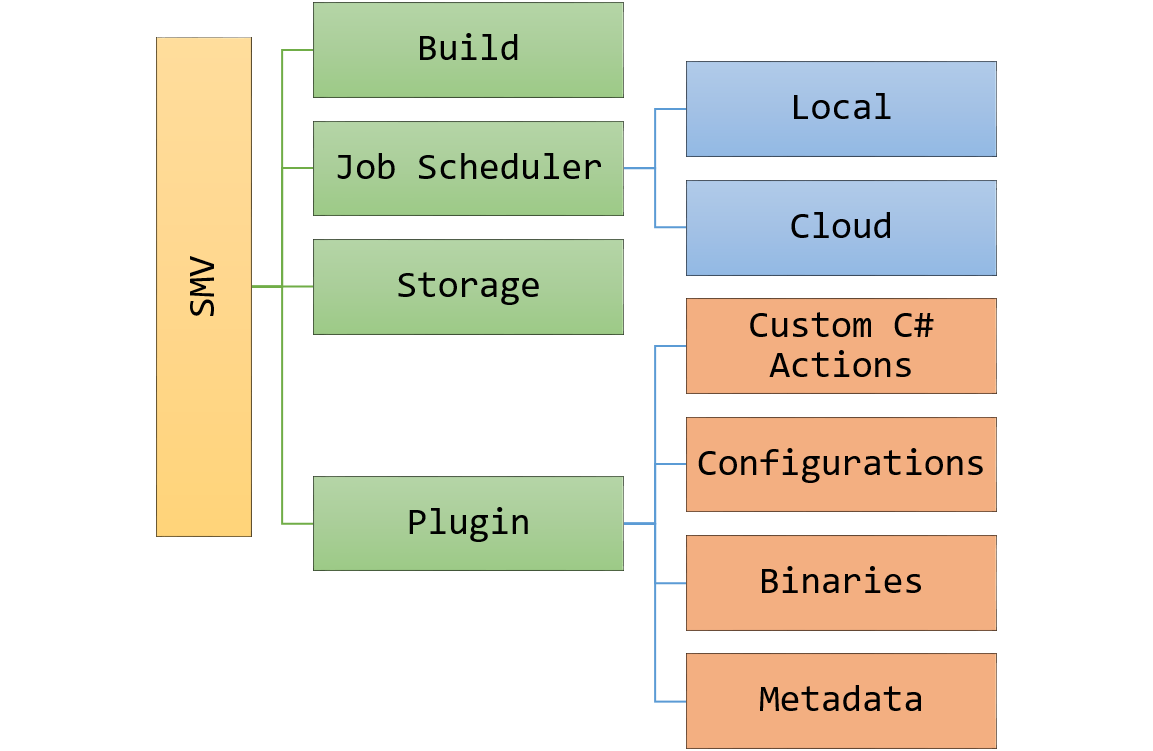}
  \caption{High level architecture of \smv{}.
    \label{fig:SMVArchitecture}}
\end{figure}

We now briefly present the architecture of the \smv{}
platform. Figure~\ref{fig:SMVArchitecture} shows the high level design
of \smv{}. The main building blocks of \smv{} are:

\begin{itemize}
\item {\em Build}. \smv{} supports several build environments, out of
  the box like MSBuild and Make. As a first step, it tries to do a
  normal build followed by an intercepted build. The intercepted build
  produces the build artifacts required by the analysis engine. For
  the intercepted build, we intercept calls to the compiler and the
  linker, which produces the rawcfg and li files, respectively. These
  are then consumed and analysed by the engine. This step is fully
  customizable -- any binary can be intercepted and then a sequence of
  events can be performed in its stead.

\item {\em Job Scheduler}. Once the build stage has successfully
  completed, IR is available to be consumed by the analysis stage. The
  IR along with the exact command to invoke the analyses are provided
  as input to the job scheduler. The job scheduler is responsible for
  scheduling the various tasks that are created in the analysis
  stage. Currently, we have two implementations: a local execution or
  a Microsoft Azure based scheduler that schedules jobs on the
  cloud. The infrastructure for performing the job on the cloud is a
  slight generalization from the \csdv{} infrastructure presented
  earlier.
  
\item {\em Storage}. Currently all jobs that are submitted to the
  cloud are archived. The archive entry includes the job inputs, exact
  command line for executing the analysis, and the analysis engine(s)
  that were used. Today, we only support storage in Microsoft Azure,
  but in the future we plan on generalizing this to support multiple
  data sources and sinks.
  
\item {\em Plugin}. \smv{} is designed with a plugin based
  architecture. Each plugin consists of the following pieces:
  
  \begin{itemize}
  \item {\em XML configuration files}. These are the tool specific XML
    files that tell \smv{} how to build, intercept, and analyze the
    module in question.
  \item Optional C\# code that implements the SMVPlugin interface which
    can be used to perform pre-build, post-build, pre-analysis,
    post-analysis, actions, parse specialized command line arguments
    etc.
  \item The tool binaries and configuration files and other meta-data
    itself. For example, if we have an engine that requires some data
    files, the plugin would contain the binaries and the data files.
  \end{itemize}
  
\end{itemize}

\section{Related Work}
\label{sec:related}

In the past, parallel techniques for model checking have been explored
in great depth.~\cite{sternDillParallelMurphi} specifically focused on
parallelizing the Murphi model checker for speeding up the exploration
of states and also possibly achieving higher scalability by exploring
more states and verifying larger models. Kumar et. al. present work in
\cite{kumar2005load} that performs load balancing of parallel model
checking algorithms. Work presented in ~\cite{BarnatScalable}
investigates how to perform LTL model checking in a distributed
environment. Work presented in~\cite{holzmannSwarm} also aims at
taking advantage of availability of greater resources and computation
power.

Given the body of work in parallel model checking, and the rise of
static analysis as a more practical solution for certain problems, it
was only a matter of time before the case for static analysis in the
cloud was made~\cite{fahndrichcase}. To this
effect,~\cite{beyer2014software} presents work that ports the
\texttt{CPAChecker} to the Google App-Engine and exposes the abilities
through API as well as a web interface. To our knowledge, this work is
the closest to the work presented in this paper but has key
differences. Primarily, the work in~\cite{beyer2014software} focuses
on using the Google App-Engine in a Platform as a Service (PaaS)
setting, while our work is focused on using Microsoft Azure in a
Infrastructure as a Service (Iaas) setting. Due to this primary
difference, porting the verification technologies involved no effort
in \csdv{}, whereas~\cite{beyer2014software} had to make significant
changes to \texttt{CPAChecker} for it to function correctly using the
Google App-Engine.

Cloud based testing services have also become more practical and
popular. \cite{ciortea2010cloud9} and~\cite{candea2010automated}
present cloud based frameworks and environments for performing
automated testing.

Scan-Build~\cite{scan-build}, is a tool which is part of the
LLVM/Clang compiler codebase~\cite{lattner2008llvm}. The Scan-Build
tool is similar to the first stage of \smv{}, wherein it
systematically intercepts calls to the compiler for a Make based
build, and performs static analysis on the source files. Although
similar in theory, there are multiple differences relative to
\smv{}. \smv{} is customizable and build system
agnostic. Additionally, \smv{} is also capable of performing
interception for any binary in question, while Scan-Build only
performs interception for calls to specific compiles. Finally,
Scan-Build only performs static analysis using the Clang analyses, or
a Clang based plugin that has been developed by a user. \smv{} on the
other hand allows analysis to be performed for the entire module, not
a single file at a time, and allows the use of any analysis technology
that is available.

\section{Conclusion}
\label{sec:conclusions}

We have presented a method for parallelizing the \sdv{} verification
tool using Microsoft Azure. The architecture and implementation make
use of core concepts provided by Microsoft Azure (blobs, queues,
workers, Autoscale etc.). Using the \csdv{} implementation, we were
able to perform large scale verification of drivers and \sdv{} test
suites in a sound and consistent manner. Our results show that the
\csdv{} implementation is extremely performant and scalable. At {\em
  worst}, \csdv{} performs as well as a local verification run, and in
the {\em best} case, \csdv{} is capable of delivering extremely large
amounts of speedup. We conclude that the observed speedup is directly
proportional to the amount of verification checks that can be
submitted to the \csdv{} system. From our current experiments and
results, we observe that \csdv{} is extremely effective for large test
suites containing a lot of verification tasks, but not as effective
for single drivers. 

Currently, the \csdv{} implementation is being evaluated for
integration with the primary \sdv{} product that is shipped with the
Windows Driver Kit. The evaluation is primarily for the purpose of
exposing the \csdv{} service to driver developers, both internal and
external. As future work in this specific area, we plan on performing
more experiments with different configurations of the Autoscale
feature to identify a possible {\em sweet spot}. We also plan on
investigating the potential to make the \csdv{} client asynchronous
(remove polling) and {\em offline}, where one can schedule jobs and
exit the \csdv{} client (not be required to have the client running
continuously).

We have also presented \smv{}, a more generalized version of \sdv{}
and \csdv{} that allow for verification of any module. The technology
in \smv{} is highly configurable, portable, and extensible. Currently
\smv{} is being evaluated as a generic verification platform to be
used within Microsoft.

\section{Acknowledgments} 

We would like to thank B. Ashok, Apoorv Upreti, Vlad Levin, and Thomas
Ball for their valuable input and support of this work.

\bibliographystyle{eptcs}
\bibliography{biblio}
\end{document}